\documentclass[aps,preprint,epsfig,rotate]{revtex4}
\usepackage{graphicx}
\usepackage{bm}
\usepackage{epsfig}

\input epsf
\begin{document}
\title{Bound state properties of four-body muonic quasi-atoms}

 \author{Alexei M. Frolov}
 \email[E--mail address: ]{afrolov@uwo.ca}

 \author{David M. Wardlaw}
 \email[E--mail address: ]{dwardlaw@uwo.ca}

\affiliation{Department of Chemistry\\
 University of Western Ontario, London, Ontario N6H 5B7, Canada}

\date{\today}

\begin{abstract}

Total energies and various bound state properties are determined for the
ground states in all six four-body muonic $a^{+} b^{+} \mu^{-} e^{-}$
quasi-atoms. These quasi-atoms contain two nuclei of the hydrogen isotopes
$p^{+}, d^{+}, t^{+}$, one negatively charged muon $\mu^{-}$ and one
electron $e^{-}$. In general, each of the four-body muonic $a^{+} b^{+}
\mu^{-} e^{-}$ quasi-atoms, where $(a, b) = (p, d, t)$, can be considered as
the regular one-electron (hydrogen) atom with the complex nucleus $a^{+}
b^{+} \mu^{-}$ which has a finite number of bound states. Furthermore, all
properties of such quasi-nuclei $a^{+} b^{+} \mu^{-}$ are determined from
highly accurate computations performed for the three-body muonic ions $a^{+}
b^{+} \mu^{-}$ with the use of pure Coulomb interaction potentials between
particles. It is shown that the bound state spectra of such quasi-atoms are
similar to the spectrum of the regular hydrogen atom, but there are a few
important differences. Such differences can be used in future experiments to
improve the overall accuracy of current evaluations of various properties of
hydrogen-like systems, including the lowest-order relativistic and QED
corrections to the total energies.


\end{abstract}
\maketitle
\newpage

\section{Introduction}

Recently, by performing variational calculations of the bound states in
three-body muonic molecular ions \cite{FroWar} we have re-discovered a new
class of four-body systems which are of interest in some applications. The
existence of these neutral four-body atomic systems follows from the
analysis of actual experimental conditions which can be found in liquid
hydrogen/deuterium. Indeed, in actual experiments with muons it is hard to
imagine a positively charged muonic molecular ion, e.g., the $(pd\mu)^+$
ion, which moves as a separate and stable system in liquid
hydrogen/deuterium with density $\rho \approx 0.213$ $g \cdot cm^{-3}$.
It is clear that such a system will take an additional electron $e^{-}$ from
surrounding atoms/molecules and form the neutral muonic quasi-atom $pd\mu
e$. Later such a four-body quasi-atom $pd\mu e$ will react with a
protium/deuterium molecule and form a stable six-body quasi-molecule
$(pd\mu) p e_2$ or $(pd\mu) d e_2$. All transformations of muonic four-body
quasi-atoms and six-body quasi-molecules occur on a time scale of $\tau
\approx 1 \cdot 10^{-10}$ $sec$ which is significantly shorter than the muon
life time $\tau_{\mu} \approx 2 \cdot 10^{-6}$ $sec$ and/or the reaction
time for the nuclear $(p,d)-$reaction. This means that the muonic
quasi-atoms $pd\mu e$, $pd\mu e$, $dt\mu e$, etc, and muonic quasi-molecules
$(pd\mu) p e_2$ and $(pd\mu) d e_2$ can be observed in actual experiments
and it is therefore interesting to investigate the properties of such systems.
In this study we determine the bound state properties of the four-body $a b
\mu e$ quasi-atoms and discuss some interesting experiments for these
systems.

For concisness of presentation, we confine attention to the $p d \mu e$ 
quasi-atom; extention to the other five hydrogen isotopes is straightforward. 
In atomic units $\hbar = 1, m_e = 1, e = 1$ the Hamiltonian of the four-body
$pd\mu e$ quasi-atom is:
\begin{eqnarray}
 H = -\frac{1}{2 m_{p}} \Delta_{1} -\frac{1}{2 m_{d}} \Delta_{2}
     -\frac{1}{2 m_{\mu}} \Delta_{3} -\frac{1}{2} \Delta_{4}
     + \frac{1}{r_{12}} - \frac{1}{r_{13}} - \frac{1}{r_{14}}
     - \frac{1}{r_{23}} - \frac{1}{r_{24}} + \frac{1}{r_{34}} \label{eq1}
\end{eqnarray}
where the notation 1 designates the protium nucleus ($p$ or ${}^1$H), the
notation 2 (or +) means the deuterium ($d$) nucleus, while 3 and 4 stand
for the negatively charged muon $\mu^{-}$ and electron $e^{-}$,
respectively. This system of notations will be used everywhere below in this
study. Note that the three heavy particles $p^{+}, d^{+}$ and $\mu^{-}$ in
this system have very large masses in comparison with the electron mass
$m_e$ (in atomic units $m_e = 1$) and opposite electric charges. Therefore,
we can expect that inside of the $pd\mu e$ quasi-atom these three heavy
particles ($p^{+}, d^{+}$ and $\mu^{-}$) will form a separate three-body
cluster $pd\mu$ which is spatially compact and has a positive electric
charge +1. The fourth particle (electron $e^{-}$) moves at very large
distances from this central, heavy  cluster. If $a_0$ is the Bohr radius,
then the radius of the electron orbit in the $pd\mu e$ quasi-atom is $R_e
\approx a_0$, while the spatial radius of the central heavy cluster is
$R_{\mu} \approx \Bigl( \frac{m_e}{m_{\mu}}\Bigr) a_0 \approx \Bigl(
\frac{1}{206.768262} \Bigr) a_0 \ll a_0$. Analogous relations between radii
of the electron and muonic orbits can be found in other four-body
quasi-atoms $a b \mu e$. In fact, all bound state properties in this
family of quasi-atoms $a b \mu e$ can be separated into two different
groups: electronic properties and muonic properties. Sometimes, it is
convenient to split the muonic properties into two additional subgroups:
muonic and nuclear properties. The reason for such a separation is obvious,
since each of the nuclear masses $M_p, M_d, M_t$ is substantially larger
than the muon mass $m_{\mu} = 206.768262 m_e$.

\section{Total energies}

Based on the cluster structure of the four-body quasi-atom $pd\mu e$ we can
predict that this quasi-atom is a bound four-body system which is very
similar to the regular hydrogen atom. Moreover, its binding energy must be 
very close to the total energy of the one-electron
hydrogen atom. For instance, the binding energy of the ground $1S(L =
0)-$state in the $pd\mu e$ atom must be close to -0.5 $a.u.$, while the
analogous energy of the $2S(L = 0)-$state (electron state) in the $pd\mu e$
quasi-atom must be close to -0.125 $a.u.$, etc. In general, the total
energy of the bound state with the principal quantum number $n$ in the $pd
\mu e$ quasi-atom is approximately equal to
\begin{equation}
 E \approx E(pd\mu) + E({\rm H}) \approx - 106.012 527 069 515 844
 - \frac{1}{2 n^2} \; \; \; \; a.u. \label{eq2}
\end{equation}
where the value - 106.012 527 069 515 844 $a.u.$ is the best-to date energy
of the ground state in the $pd\mu$ ion \cite{FroWar}. Thus, the ground 
state ($n$ = 1) in this four-body quasi-atom $pd \mu e$ has the total 
energy $E \approx$ -106.512 527 069 515 844(2) $a.u.$ \cite{FroWar}. 
Analogous relations can be found for other $a b \mu e$ quasi-atoms. The 
formula, Eq.(\ref{eq2}), is valid in those cases when we can neglect all 
electron-muon, electron-deuteron and electron-proton correations. In 
reality, such correllations contribute to the total energy and the value 
of $E$ given by Eq.(\ref{eq2}) must be corrected.

It is clear that the total energy $E$ depends upon particle masses. In this 
study we shall assume that $m_p$ = 1836.152701 $m_e$, $m_d$ = 3670.483014 
$m_e$, $m_t$ = 5496.92158 $m_e$ and $m_{\mu}$ = 206.768262 $m_e$. These 
masses are often used in accurate computations of the muonic molecular 
ions. With these masses the total energy of the four-body quasi-atom $pd 
\mu e$ in its ground state is $E \approx$ -106.512 439 401 $a.u.$, i.e. it 
is slightly above the value -106.512 527 069 52 $a.u.$ mentioned above. It 
is clear that the non-relativistic energy $E$ from Eq.(\ref{eq2}) is only 
an approximation to the exact total energy of the $p d\mu e$ quasi-atom, 
since this value of $E$ does not include all contributions from the 
lowest-order relativistic and QED corrections. A large number of other 
corrections, e.g., the finite-size corrections, corrections on nuclear 
interactions, etc, to the total energy have been ignored as well.

The total energy $E$ given by Eq.(\ref{eq2}) corresponds to the case when
the central quasi-nucleus (or heavy three-body cluster) $pd\mu$ is in its
ground $S(L = 0)-$state. However, this three-body quasi-nucleus $pd\mu$
can also be detected in its bound $P(L = 1)-$state with the total energy
-101.453 777 548 914 $a.u.$ \cite{FroWar}. In this case the total energy of
the $pd\mu e$ quasi-atom with the central nucleus in its bound $P(L =
1)-$state is $\approx$ -101.953 777 548 914 $a.u.$, i.e. it is substantially
different from the value $E \approx$ -106.512 439 401 $a.u.$ mentioned above
for the ground state and the main difference arises from the energy of the
central quasi-nucleus $pd\mu$.

As is well known the six muonic molecular ions $pp\mu, pd\mu, pt\mu, dd\mu,
dt\mu$ and $tt\mu$ have 22 bound states (see, e.g., \cite{FroWar} and
references therein). These are the $S(L = 0)-, P(L = 1)-, D(L = 2)-$ and
$F(L = 3)-$states, where the notation $L$ designates the total angular
momentum of this three-body ion $ab\mu$, where $(a, b) = (p, d, t)$. There 
are nine (bound) $S(L = 0)-$states, nine $P(L = 1)-$states, three
$D(L = 2)-$states and one $F(L = 3)-$state. The four-body muonic quasi-atoms
which contain two nuclei of hydrogen isotopes can be considered as a family
of similar one-electron (or hydrogen-like) atoms. One of the six muonic
molecular ions $pp\mu, pd\mu, pt\mu, dd\mu, dt\mu$ and $tt\mu$ plays the
role of the nucleus in each of these atoms. In general, such a nucleus can
be either in the ground $S(L = 0)-$state, or in one of its `rotationally'
and/or `vibrationally' excited states. Here to designate the
excited states in the $ab\mu$ ion we use the $(L,\nu)-$notation, where $L$
denotes the rotational state, while $\nu$ stands for the vibrational state.

All bound states properties, including the lowest order relativistic
and QED corrections, determined for such hydrogen-like quasi-atoms
depend substantially upon the hydrogen-isotope composition and excitation of
the central three-body `nucleus' $ab\mu$. The spectrum of the excited states
is very well known for each of the six muonic molecular ions $ab\mu$.
Therefore, the $ab\mu e$ quasi-atoms with different hydrogen isotopes $a =
(p, d, t)$ and $b = (p, d, t)$ can be considered as model hydrogen-like
atoms. The `nuclear' spectra in such atoms are known to very high accuracy.
This allows one to consider possible interactions between `atomic' and
`nuclear' bound states in the $ab\mu e$ quasi-atoms. Note again that all
`nuclear' properties of the $ab\mu$ quasi-nucleus can be obtained from
highly accurate Coulomb calculations performed for three-body systems.

In general, the electron motion in the $a b \mu e$ quasi-atom is well
separated from the motion of the three heavy particles $a^{+} b^{+}
\mu^{-}$. For instance, the expectation values of the kinetic energies of
the electron and muon in the $a b \mu e$ quasi-atom differ by a factor of 50
- 100. However, as is well known (see, e.g., \cite{MS}, \cite{BhaDra} and
references therein) the `vibrationally' excited $P(L = 1)-$states (or
(1,1)-states) in the $dd\mu$ and $dt\mu$ three-body ions are very weakly
bound. The binding energies of the three-body ions $dd\mu$ and $dt\mu$ in
their excited (1,1)-states are $\approx$ -1.9749881 $eV$ and $\approx$
-0.6603387 $eV$, respectively \cite{FroWar}. These `nuclear' binding
energies are comparable with the corresponding atomic energies. In such
cases one can certainly observe a strong interference between the `nuclear'
and `atomic' bound states. In reality, for bound states with very weakly
bound nuclei (or quasi-nuclei) we cannot discuss their nuclear and electron
spectra separately. The analysis of atomic systems with weakly bound nuclei
is very complex, but its results are of great interest in a number of
applications.

\section{Variational wave functions}

To determine the total energies and bound state properties one needs to
construct approximate wave functions for the four-body quasi-atoms. In this
study these wave functions are approximated by the variational expansion
written in the basis of the six-dimensional (or four-body) gaussoids of
relative (or inter-particle) scalar coordinates $r_{ij}$. This variational
expansion was originally proposed 30 years ago in \cite{KT} for accurate
variational calculations of few-nucleon nuclei and $\Lambda-$nuclei. For
the bound $S(L = 0)-$states the variational anzatz of fully correlated
six-dimensional (or four-body) gaussoids is written in the form \cite{KT}
\begin{eqnarray}
 \Psi_{L=0} = (1 + \delta_{ab} \epsilon_{ab} {\cal P}_{ab}) \sum_{k=1}^N
 C_k \cdot exp( -\alpha^{(k)}_{12} r^{2}_{12} -\alpha^{(k)}_{13} r^{2}_{13}
 -\alpha^{(k)}_{23} r^{2}_{23} -\alpha^{(k)}_{14} r^{2}_{14}
 -\alpha^{(k)}_{24} r^{2}_{24} -\alpha^{(k)}_{34} r^{2}_{34}) \label{Gaus}
\end{eqnarray}
where $C_k$ are the linear coefficients (or linear variational parameters),
while $\alpha^{(k)}_{ij}$ are the optimized non-linear parameters. The
notation $\epsilon_{ab} {\cal P}_{ab}$ means the appropriate symmetrizer (or
antisymmetrizer), i.e. a projection operator which produces the wave
function with the correct permutation symmetry in those cases when $a = b$.
This case is designated in Eq.(\ref{Gaus}) with the use of the
delta-function. The operator ${\cal P}_{ab}$ is the pair-permutation
operator for all coordinates, i.e. for the spatial, spin, iso-spin, etc,
coordinates.

By using some effective strategies for optimization of the non-linear
parameters $\alpha^{(k)}_{ij}$ in Eq.(\ref{Gaus}) one can obtain very
accurate variational wave functions with relatively small number of terms $N
\approx$ 400 - 600 in Eq.(\ref{Gaus}). The generalization of the variational
expansion, Eq.(\ref{Gaus}), to the bound states $L \ge 1$ is
straightforward. It is clear that for the quasi-atoms $a b \mu e$ with the
`rotationally' excited `nuclei' $a b \mu$ the minimal number of terms in
Eq.(\ref{Gaus}) must be larger to achieve better overall accuracy. However,
in this study we shall not discuss numerical computations of bound states
with $L \ge 1$ and restrict ourselves to the analysis of the bound $S(L =
0)-$states.

\section{Binding energies and other bound state properties}

As we have shown above the total energy of the $pd\mu e$ atom is
approximately equal to the sum of the total energies of the hydrogen atom
and $pd\mu$ ion, i.e. $E(pd\mu e) \approx E(pd\mu) + E({\rm H}; n\ell)$,
where $n$ and $\ell$ are the principal and angular quantum numbers of the
hydrogen atom. The corresponding binding energy is the difference between
this value and the total energy of the three-body $pd\mu$ ion. This means
that the binding energy of the ground state of the four-body atom $pd\mu e$
approximately equals to the total energy of the ground state of the hydrogen
atom which equals -0.5 $a.u.$ In other words, the total energy of the ground
state in the four-body $pd\mu e$ system approximately equals the sum of
total energies of the three-body muonic molecular ion $pd\mu$ and the ground
state energy of the hydrogen atom with the infinitely heavy nucleus, i.e. 
-0.5 $a.u.$ The use of the finite masses for different hydrogen isotopes $p, 
d, t$ and muon $\mu$ slightly decreases the absolute value of the total 
energy of the $p d \mu e$ system. The corresponding binding energies of all 
six $a b \mu e$ quasi-atoms in their ground states $\varepsilon_H$ can be 
found in Table I. These values have been determined with the use of our data 
from Table I and \cite{FroWar} and the following formula
\begin{eqnarray}
  \varepsilon_H = E(a b \mu e) - E(a b \mu)
\end{eqnarray}
As follows from Table I these (electron) binding energies are realy close to 
the expected value -0.5 a.u. A very accurate evaluation of $\varepsilon_H$ is
given by the formula
\begin{eqnarray}
  \varepsilon_H \approx \varepsilon_A = - \frac{0.5}{1 + \frac{1}{m_a + m_b 
  + m_{\mu}}}
\end{eqnarray}
The values of $\varepsilon_H$ and $\varepsilon_A$ can also be found in Table 
I. They correspond to the total energies obtained with the use of $N$ = 600
basis wave functions in Eq.(\ref{Gaus}).

By using our total energies computed with the use of $N$ = 400 and $N$ = 600
basis functions we can extrapolate our results to the infinite number of 
basis functions. The corresponding energy $E(N = \infty)$ must be closer to 
the actual total energy than each of the $E(N = 400)$ and $E(N = 600)$ 
energies. In general, the following extrapolation formula is often used for
this purpose
\begin{eqnarray}
  E(N_i) = E(N = \infty) + \frac{A}{N^{\gamma}_i} \label{extr}
\end{eqnarray}
From a series of calculations performed for four-body muonic quasi-atoms we
have found that for such systems the parameter $\gamma$ in Eq.(\ref{extr})
varies between 3.5 and 4. Below, we shall assume that $\gamma = 3.5$ in
Eq.(\ref{extr}). Now the formula, Eq.(\ref{extr}), can be applied to all
four-body muonic quasi-atoms mentioned in Table I. It sould be mentioned, 
however, that this extrapolation formula can be applied only in those cases,
when the internal (or non-linear) parameters of this method are not changed 
(or not varied). In variational expansion, Eq.(\ref{Gaus}), the nonlinear 
parameters are always varied to produce the results of good numerical accuracy. 
Therefore, the parameters $E(N = \infty)$ and $A$ becomes $N-$dependent. This
means that by using different values of $E(N = 400)$ and/or $E(N = 600)$ one 
finds a number of different $E(N = \infty)$ values. Formally, we have a 
distribution of $E(N = \infty)$ values which can be written in the form 
$E(N = \infty) = \tilde{E}(N = \infty) \pm \Delta$, where $\Delta$ is the
corresponding uncertainty. The values of $\tilde{E}(N = \infty)$ and $\Delta$
can be found in Table I for each muonic quasi-atom.

Other bound state properties of the $a b \mu e$ quasi-atoms computed with
our approximate wave functions (see Table II), e.g., the $\langle r_{ij} 
\rangle, \langle r^2_{ij} \rangle, \langle \delta({\bf r}_{ij}) \rangle$ 
expectation values, coincide well either with the known properties of the 
hydrogen atom (in those cases, when one of the indexes $i$ or $j$ designates 
the electron), or with the bound state properties known for the three-body 
muonic molecular ion $ab\mu$ (see, e.g., \cite{FroWar} and references 
therein). This uniformly indicates that each of the $a b \mu e$ quasi-atoms 
has the two-shell cluster structure. The electron moves at substantial 
(atomic) distance from compact central cluster $a b \mu$. This central 
cluster $a b \mu$ has a structure which is similar to a molecular (or 
two-center) structure. It is different from expected `pure nuclear' (or 
one-center) structure. The effective spatial radius of the central cluster 
is in $m_{\mu} \approx$ 206.768 times smaller than the radius of the 
electron orbit. All these conclusions directly follow from Table II.  

Note that the set of operators included in Table II is not exaustive and 
was selected for illustrative purposes. Most of the bound state properties 
from Table II have been determined to relatively high accuracy which can be
be estimated by comparing the corresponding expectation values computed 
with the use of 400 and 600 basis functions. The overall accuracy of the 
$\langle r_{ij}^{-1} \rangle$ expectation values (7 - 8 stable decimal 
digits) can be estimated by using the virial theorem. The expectation 
values  $\langle r_{ij}^{n} \rangle$, where $n = -2, 1, 2$ have close
overall accuracy. Anlogously, the number of correct decimal digits in the 
expectation values of the partial kinetic energies (or single-particle 
kinetic energies) $\langle -\frac12 \nabla^{2}_i \rangle$ can also be 
evaluated with the use of virial theorem. The expectation values of the 
electron-nuclear delta-functions contain $\approx$ 6 stable decimal digits, 
while the muon-nuclear and electron-muon delta-functions are slightly less 
accurate ($\approx$ 4 - 5 stable decimal digits). The expectation values of 
the delta-functions between two nuclei of hydrogen isotopes include only 1 
- 2 accurate decimal digits. These expectation values must be improved in 
future calculations. The computed expectation values can be compared directly 
with the expectation values obtained earlier for the three-particle $pd\mu, 
pt\mu$ and $dt\mu$ ions and for the one-electron hydrogen atom. In general, 
such a comparison is of great interest, since it allows to compare 
directly the overall qualities of the different variational wave functions. 
On the other hand, it is very interesting to find electronic properties of 
the four-body $ab\mu e$ quasi-atoms which are substantially different from 
the known properties of the hydrogen atom(s).

The expectation values of different bound state properties determined to
high numerical accuracy allow one to estimate a large number of fundamental
atomic properties which can be directly measured in actual experiments. Here
by `fundamental property' we shall understand some combination of atomic
expectation values which determine the results of direct experimental
observations. In other words, any fundamental property leads to some
experimental effects. Below, we discuss only the two following effects: (a)
the evaluation of the field component of the total isotope shift for the $a
b \mu e$ quasi-atoms, and (b) the hyperfine structure splittings in these
quasi-atoms. As is well known (see, e.g., \cite{Sob}, \cite{Fro07} and
references therein) the field shift in atomic systems is related to the 
extended nuclear charge distribution which produces the non-Coulomb field 
at distances close to the nucleus, i.e. at $r \approx R_N \approx 10^{-13}$ 
$cm$ (= 1 $fm$) where $R_N$ is the radius of the nucleus. It is clear that 
the largest deviations between the Coulomb and actual potentials can be 
found close to the atomic nucleus, i.e. for distances $r \approx R_N 
\approx r_e \ll \Lambda \ll a_0$, where $r_e = \alpha^2 a_0$ is the 
classical electron radius and $\Lambda = \alpha a_0$ is the Compton wave
length. Here and below, $\alpha = 7.297352568 \cdot 10^{-3}$ is the fine
structure constant and $a_0 \approx 5.29177249 \cdot 10^{-11}$ $m$ is the
Bohr radius. An approximate formula for the field shift $E^{fs}_M $ in 
light atoms takes the form \cite{Fro11}
\begin{eqnarray}
  E^{fs}_M = \frac{8 \pi}{5} Q \alpha^4 \xi \Bigl(\frac{R_N}{r_e}\Bigr)^2
\end{eqnarray}
where $Q$ is the nuclear charge and $\xi$ is an additional factor which is 
uniformly related to the charge distribution in the nucleus \cite{Fro11}. 
An essentially equivalent formula was given long ago by Cooper and Henley
\cite{Cooper}. Now note that all `nuclei' in $a b \mu e$ quasi-atoms are 
three-body muonic molecular ions $a b \mu$ with the spatial radius $a_{\mu} 
\approx a_0 / m_{\mu} \gg r_e = \alpha^2 a_0$. Therefore, we can predict
that the field component of the total isotope shift in the $a b \mu e$
quasi-atoms must be substantially larger (10,000 times larger) than for 
the usual light atoms. Furthermore, this component of the total isotope 
shift will be $a-$ and $b-$ dependent, since the spatial radius of the 
$a b \mu e$ quasi-atom depends upon the two nuclei of the hydrogen 
isotopes $a$ and $b$.

Another interesting problem is to predict the hyperfine structure splitting
of the four-body $pd\mu e$ and other similar quasi-atoms. Later the computed
hyperfine structure splitting must be compared with the corresponding
experimental results. For instance, consider the hyperfine structure
splitting of the four-body $pd\mu e$ quasi-atom. The central quasi-particle,
i.e. the $pd\mu$ ion, has two bound states with different angular momenta.
These states are also bound in the analogous $pt\mu$ and $pp\mu$ ions. In
each of the $dd\mu$ and $dt\mu$ ions one finds five bound states: two $S(L =
0)-$states, two $P(L = 1)-$states and one $D(L = 2)-$state. The heaviest
$tt\mu$ ion has six bound states: two $S(L = 0)-$states, two $P(L =
1)-$states, one $D(L = 2)-$state and $F(L = 3)-$state. We can define the
angular moment $L$ of the central `nucleus' $ab\mu$ and the magnetic moment
${\bf S}$ (or `spin', for short) associated with $L$. The `nuclear' spin
${\bf S}$ of the $ab\mu$ quasi-nucleus is combined with the electron spin
${\bf s}_e$ and this produces the hyperfine structure splitting of the
four-body $ab\mu e$ quasi-atom. By determining the expectation values of
the corresponding delta-functions and a few other properties one can
evaluate the hyperfine structure splitting in all $S(L = 0)-$states of the
six four-body $ab \mu e$ systems, where $(a, b) = (p, d, t)$. Accurate
numerical evaluation of the hyperfine splitting in the $ab \mu e$ systems
will be our goal in an upcoming study. The computed values of the 
hyperfine structure splittings must be compared with the actual experimental 
values. Analysis of the hyperfine structure splitting for the rotationally 
excited states is more complicated.

In some sense the four-body muonic quasi-atoms are the two-shell atomic 
systems which are similar to the helium-muonic atoms discussed in
\cite{PRA1978}, \cite{PRA1980} and \cite{Fro02} (see also references 
therein). In particular, these four-body muonic quasi-atoms, e.g., $p d 
\mu e$, have a very compact central cluster ($pd\mu$) and one electron
moving in the electric field of this cluster. The fundamental difference
between the $p d \mu e$ quasi-atom and ${}^4$He$\mu e$ atom follows from
the different nature of their central clusters. A very heavy nucleus 
${}^4$He (or ${}^3$He) is essentially the center of the ${}^4$He$\mu e$ 
atom, while the two positively charged particles $p$ and $d$ in the 
$pd\mu$ quasi-nucleus are not bound to each other without the muon. Briefly,
we can say that such a central cluster $pd\mu$ has a `molecular' 
structure.  
 
\section{Conclusions}

We have discussed the bound state spectra and properties of the muonic
four-body quasi-atoms $a b \mu e$, where $(a,b) = (p, d, t)$. The energy
spectra of such four-body quasi-atoms are of interest in some applications.
Briefly, an arbitrary bound state in the four-body muonic quasi-atom $a b
\mu e$ is represented as motion of the negatively charged electron
$e^{-}$ in the field of a compact `central cluster' $a b \mu$ which has
positive electric charge +1. The spatial radius of the central cluster in
the $a b \mu e$ quasi-atom is in $\approx 206.77$ times smaller than the
actual radius of the electron orbit $r_e \approx a_0$, where $a_0$ is the
Bohr (atomic) radius. To the best of our knowledgr this study is the first 
analysis of the bound state properties of the four-body muonic quasi-atoms 
$a b \mu e$. In general, the study of bound state spectra in four-body 
muonic quasi-atoms $ab \mu e$ may lead to some important results and 
conclusions which can also be useful in applications to other atomic and 
quasi-atomic systems. In reality, for each four-body muonic quasi-atoms $a 
b \mu e$ we know the bound state spectra of its central quasi-nucleus $a b 
\mu$. Moreover, the probabilities of all possible electromagnetic 
transitions in this quasi-nucleus ($ab \mu$) can be determined to very 
good accuracy. This simplifies the analysis of many related phenomena, 
including internal radiative transitions, in the four-body quasi-atom $a b 
\mu e$.

There are a number of related problems which are of special interest in
various applications. In particular, accurate computation of some bound 
state properties for the four-body muonic quasi-atoms $ab \mu e$, including 
the lowest order relativistic and QED corrections to the total and binding 
energies. There is great interest in determining such corrections from 
direct calculations and comparing them with the results of approximate 
evaluations based on the one-electron model of the $a b \mu e$ quasi-atom. 
In general, the theoretical and experimental analysis of four-body 
quasi-atoms $ab \mu e$ is a reach area of study. Analysis and solutions of 
this problem will certainly lead to a substantial improvement of our current 
knowledge about the bound state spectra in four-body systems with arbitrary 
particle masses.

\newpage
  \begin{table}[tbp]
   \caption{The total non-relativistic energies $E$ of the ground
            states of the six four-body muonic quasi-atoms $a b
            \mu e$ (in atomic units). $N$ designates the total
            number of basis functions used in Eq.(3).}
     \begin{center}
     \scalebox{0.95}{%
     \begin{tabular}{cccc}
      \hline\hline
 $E(N)$ & $p d \mu e$ & $p t \mu e$ & $d t \mu e$ \\
     \hline
 $E(N = 400)$    & -106.51225045   & -107.99443352 & -111.86418333 \\

 $E(N = 600)$    & -106.51230138   & -107.99446758 & -111.86419175 \\

  $\tilde{E}(N = \infty)(\Delta)$  & -106.5125(2) & -107.9947(2)  & -111.8645(2) \\
     \hline 
 $\varepsilon_H$ & -0.49977431 & -0.49976497 & -0.49984433 \\

 $\varepsilon_A$ & -0.49991250 & -0.49993369 & -0.49994667 \\
     \hline\hline
 $E(N)$ & $p p \mu e$ & $d d \mu e$ & $t t \mu e$ \\
     \hline
 $E(N = 400)$  & -102.72329207 & -110.31679566  & -113.47261012 \\

 $E(N = 600)$  & -102.72330204 & -110.31680450  & -113.47261845 \\

 $\tilde{E}(N = \infty)(\Delta)$ & -102.72335(3) & -110.31685(3) & -113.47265(3) \\
    \hline
 $\varepsilon_H$ & -0.49979846 & -0.49987811 & -0.49976942 \\

 $\varepsilon_A$ & -0.49987114 & -0.49993376 & -0.49995536 \\
    \hline\hline
  \end{tabular}}
  \end{center}
  \end{table}
   \begin{table}[tbp]
   \caption{The expectation values $\langle X_{ij} \rangle$ in atomic
            units ($m_{e} = 1, \hbar = 1, e = 1$) of some properties
            for the ground states of the $p d \mu e, p d \mu e$ and 
            $d t \mu e$ ions. Below, the notations 1 and 2 designate 
            the two heavy hydrogen nuclei, while 3 stands for the 
            negatively charged muon and 4 denotes the electron.}
     \begin{center}
     \scalebox{0.95}{%
     \begin{tabular}{cccccccc}
      \hline\hline
 $\langle X_{ij} \rangle$ & $p d \mu e$ & $p t \mu e$ & $d t \mu e$ & $\langle X_{ij}
 \rangle$ & $p d \mu e$ & $p t \mu e$ & $d t \mu e$ \\
     \hline
 $\langle r_{12}^{-2} \rangle$ & 6844.24 &  7006.69 & 7860.90 & $\langle r_{12}^{-1} \rangle$ & 76.31734 & 77.53007 & 83.49840 \\

 $\langle r_{13}^{-2} \rangle$ & 38979.3 & 37946.9  & 48009.2 & $\langle r_{13}^{-1} \rangle$ & 132.5684 & 130.9652 & 149.4313 \\
 
 $\langle r_{23}^{-2} \rangle$ & 52479.2 & 56036.3  & 52600.6 & $\langle r_{23}^{-1} \rangle$ & 155.7732 & 161.5537 & 156.7953 \\

 $\langle r_{14}^{-2} \rangle$ & 1.99803 & 1.99926  & 1.99821 & $\langle r_{14}^{-1} \rangle$ & 0.999936 & 1.000105 & 0.999971 \\

 $\langle r_{24}^{-2} \rangle$ & 1.99773 & 1.99894  & 1.99805 & $\langle r_{24}^{-1} \rangle$ & 0.999946 & 1.000120 & 0.999977 \\

 $\langle r_{34}^{-2} \rangle$ & 1.99637 & 1.99740  & 1.99714 & $\langle r_{34}^{-1} \rangle$ & 0.999893 & 1.000065 & 0.999932 \\
      \hline
 $\langle r_{12} \rangle$ & 0.0149961 & 0.0146856 & 0.0132898 & $\langle r_{12}^{2} \rangle$ & 0.253289$\cdot 10^{-3}$ & 0.242031$\cdot 10^{-3}$ & 0.193839$\cdot 10^{-3}$ \\

 $\langle r_{13} \rangle$ & 0.0118563 & 0.0119035 & 0.0102429 & $\langle r_{13}^{2} \rangle$ & 0.187901$\cdot 10^{-3}$ & 0.187857$\cdot 10^{-3}$ & 0.137575$\cdot 10^{-3}$ \\
 
 $\langle r_{23} \rangle$ & 0.0100972 & 0.0096824 & 0.0097874 & $\langle r_{23}^{2} \rangle$ & 0.137921$\cdot 10^{-3}$ & 0.126410$\cdot 10^{-3}$ & 0.126237$\cdot 10^{-3}$ \\

 $\langle r_{14} \rangle$ & 1.498820  & 1.498714 & 1.499141 & $\langle r_{14}^{2} \rangle$ & 2.990077 & 2.989908 & 2.992869 \\

 $\langle r_{24} \rangle$ & 1.498797  & 1.498686 & 1.499141 & $\langle r_{24}^{2} \rangle$ & 2.989995 & 2.989814 & 2.992831 \\

 $\langle r_{34} \rangle$ & 1.498817  & 1.498708 & 1.499158 & $\langle r_{34}^{2} \rangle$ & 2.990058 & 2.989881 & 2.992882 \\
      \hline
 $\langle -\frac12 \nabla^{2}_1 \rangle$ & 11997.364 & 12048.584 & 16719.815 & $\langle \delta_{12} \rangle$ & 168.0 & 87.2 & 9.71 \\

 $\langle -\frac12 \nabla^{2}_2 \rangle$ & 15710.188 & 17057.831 & 18036.979 & $\langle \delta_{13} \rangle$ & 1.0352$\cdot 10^{6}$ & 9.9989$\cdot 10^{5}$ & 
 1.3589$\cdot 10^{6}$ \\

 $\langle -\frac12 \nabla^{2}_3 \rangle$ & 19683.906 & 20227.987 & 21406.215 & $\langle \delta_{23} \rangle$ & 1.5244$\cdot 10^{6}$ & 1.6635$\cdot 10^{6}$ & 
 1.5331$\cdot 10^{6}$ \\

 $\langle -\frac12 \nabla^{2}_4 \rangle$ & 0.5000089 & 0.50017715 & 0.50002151 & $\langle \delta_{14} \rangle$ & 0.30981 & 0.31007 & 0.31145 \\

 $\langle \delta_{24} \rangle$ & 0.30838 & 0.30849 & 0.31041 & $\langle \delta_{34} \rangle$ & 0.30650 & 0.30729 & 0.30895 \\
      \hline\hline
  \end{tabular}}
  \end{center}
  \end{table}
\end{document}